\documentclass{article}

\usepackage{arxiv}

\usepackage[utf8]{inputenc} 
\usepackage[T1]{fontenc}    
\usepackage{hyperref}       
\usepackage{url}            
\usepackage{booktabs}       
\usepackage{amsfonts}       
\usepackage{nicefrac}       
\usepackage{microtype}      
\usepackage{lipsum}
\usepackage{amsmath}
\usepackage{graphicx}
\usepackage{enumerate}
\usepackage{lineno}
\usepackage{setspace}
\title{Collective decision-making under changing social environments among agents adapted to sparse connectivity}

\author{
  Richard P. Mann \\
  Department of Statistics, School of Mathematics, University of Leeds\\
  \texttt{r.p.mann@leeds.ac.uk}
}

\begin{document}
\maketitle


\begin{abstract}
Humans and other animals often follow the decisions made by others because these are indicative of the quality of possible choices, resulting in `social response rules': observed relationships between the probability that an agent will make a specific choice and the decisions other individuals have made. The form of social responses can be understood by considering the behaviour of rational agents that seek to maximise their expected utility using both social and private information. Previous derivations of social responses assume that agents observe all others within a group, but real interaction networks are often characterised by sparse connectivity. Here I analyse the observable behaviour of rational agents that attend to the decisions made by a subset of others in the group. This reveals an adaptive strategy in sparsely-connected networks based on highly-simplified social information: the difference in the observed number of agents choosing each option. Where agents employ this strategy, collective outcomes and decision-making efficacy are controlled by the social connectivity at the time of the decision, rather than that to which the agents are accustomed, providing an important caveat for sociality observed in the laboratory and suggesting a basis for the social dynamics of highly-connected online communities.
\end{abstract}


\section*{Introduction}
Living in groups provides many advantages to social animals \cite{krause2002liv}, including the opportunity to make use of `social information' to improve the quality of decision-making \cite{ward2011faa,wolf2013accurate}. Choices made by others are often a useful indicator of which option is best for oneself; if one trusts that others are acting rationally \cite{mann2018cdm}, and that they have similar preferences to oneself \cite{mann2020collective}, then their decisions provide valuable clues about the quality of different options. However, when others are also attending to each other this can cause information cascades \cite{bikhchandani1992theory, anderson1997information}, rendering all but a few decisions meaningless as sources of information. A crucial factor likely to determine how individuals learn from observing each other is therefore the social connectivity of the group: who each individual can observe and respond to. Previous research has investigated how rational agents make use of social information \cite{perez2011cab, arganda2013acr, mann2018cdm, mann2020collective, karamched2020heterogeneity} but these studies have focused on groups in which decisions constitute `common knowledge' \cite{aumann1976agreeing}, whereas in real groups and societies agents typically observe only a subset of other individuals. For example, visual occlusion by other individuals can restrict fish to observing a small subset of other group members \cite{strandburg2013vsn, rosenthal2015revealing, poel2021spatial}; social relationships in bird populations may form a sparse but connected network across broad geographical regions \cite{psorakis2015inferring}; and historically most human societies were characterised by sparse social connections relative to the total population \cite{dunbar1998social, apicella2012social}. As such, evolved social behaviour is likely to be explicable as an adaptive response to a social environment characterised by sparse connectivity. 

Understanding how the structure of the social environment shapes the way that individuals interact is of critical importance due to rapid changes to this structure in both human and animal populations \cite{bak2021stewardship, fisher2021anticipated}. One effect of large scale loss of animal abundance \cite{ceballos2020vertebrates, wagner2021insect} is to reduce the number and diversity interactions between con- and hetero-specifics. How animals individually and collectively respond to this social change will be a key determinant of their robustness to climate change and habitat destruction. Human societies have also been subjected to waves of dramatic change in the social landscape, the most recent of which is the rise of social media and ubiquitous, instant global connectivity. Each step change in communications technology has had profound social consequences \cite{mcquail1977influence, eisenstein1980printing}; to understand the possible implications of the latest communications revolution one must identify the factors that determine how people exchange social information and extrapolate these to determine the collective consequences of changing social connectivity \cite{fisher2021anticipated}.

Learning on social networks is a well-studied phenomena, with many theoretical treatments. The De Groot Model \cite{degroot1974reaching} and the Voter Model \cite{holley1975ergodic} present the canonical examples, and the majority of subsequent treatments have followed these by assuming that agents exhibit bounded rationality: that is, they update their opinions based on those of their social connections according to heuristic rules that are specified in advance, such as adopting the majority or average opinion of those in the social neighbourhood. While this permits precise analysis of the resulting opinion dynamics, the conclusions of such models hinges on the validity of the specified update rules, which must be justified either intuitively or empirically, rather than with respect to a deeper theory of what the agents are seeking to achieve through their decisions. 

An alternative to heuristic models is to imbue agents with goals, through utility functions, and the ability to make rational decisions through statistical inference. Sequential decision-making based on Bayesian rational decision theory is a paradigm for understanding the mechanisms of collective decision-making through such inference in both human and animal groups. Within social science (particularly economics), research has primarily focused on the efficacy of individual and collective decisions by rational agents under a variety of different scenarios \cite{bikhchandani2021information}. This research usually takes agents to be rational under the strong assumption that individuals are aware of the full details of the context in which they make the decision. Under this assumption, researchers often seek to determine properties of collective decision-making in large groups ($n \rightarrow \infty$), permitting formal mathematical answers to questions such as the conditions under which information cascades will occur \cite{bikhchandani1992theory, anderson1997information} and whether the majority of agents will decide correctly \cite{bikhchandani1992theory,acemoglu2011bayesian, mossel2015strategic}. Closely related to the current study, \cite{acemoglu2011bayesian} investigated sequential decision making by individuals in social networks that determine which other agents they can observe and identified conditions under which agents either are or are not guaranteed to converge to the correct decision. In particular, they show that convergence to the correct decision is guaranteed when agents' private beliefs are unbounded (an agent may receive arbitrarily-strong private signals), but otherwise convergence depends on the network topology -- a result that is further extended by \cite{mossel2015strategic}. However, these theorems strictly pertain only to the limiting behaviour of very large social networks, and depend on the assumption that agents hold correct beliefs regarding the structure of the social world and how other agents will respond to information, an assumption that may not hold under conditions of rapid social change or displacement to a new environment.

In animal biology a greater focus has been placed on identifying the 'interaction rules' that specify the probability that a focal individual will make a specific decision conditioned on the choices made by other animals (typically conspecifics) within the group \cite{perez2011cab, arganda2013acr, mann2018cdm, mann2020collective, mann2021optimal}. Of particular interest is the goal of deriving theoretically-motivated interaction rules that closely match empirical observations from collective decision-making in animal groups, whether in the laboratory \cite{perez2011cab, arganda2013acr, miller2013bia, kadak2020} or the field \cite{farine2013}. As such, this research typically prioritises identifying what can be observed about in social decision-making over establishing theoretical bounds on decision-making performance. In pursuit of this goal, past modelling has generally considered group sizes consistent with experimental research, often below ten individuals and rarely above 100, rather than deriving results for the limiting case of very large group sizes. In this context, most models have made the assumption, often implicitly, that each individual within the group can observe all other group members and their decisions, including the order in which those decisions were made. Recent work, however, has shown how agents can respond optimally to highly-specific forms of restricted social information, such as the aggregate number of previous agents selecting each option \cite{mann2021optimal}. 

Of special interest from a biological perspective is the question of how variation in the observable behaviour of individuals and groups can be explained through variation in context such as the quality of environmental information \cite{mann2018cdm}, degree of alignment between individual preferences \cite{mann2020collective} or changes to the overall risk level of the choices available \cite{perez2017adversity}. A distinguishing feature of the biological approach has been the consideration when investigating the effect of changing these model parameters that agents may remain adapted to a previously established context \cite{perez2017adversity, mann2018cdm, mann2020collective}, with particular consideration as to how animals are likely to behave in the laboratory as opposed to in a the natural environment \cite{mann2018cdm}.

In this paper I take a primarily biological approach to the sequential decision-making problem, characterising the observable features of the interactions between agents who act rationally based on expectations originating from their adaptation to habitual conditions. I extend previous biologically-motivated models \cite{mann2018cdm, mann2020collective, mann2021optimal} by considering agents that live on a random network, such that each individual can only observe the decisions made by a limited number of other group members: that subset of the group to which it is connected. I identify the optimal decision-making rules for agents inhabiting such networks, and in particular I focus on the form of those rules in the case where the social network is sparse, corresponding to the biologically-relevant examples given above. In this context I explore how the form of the resulting interaction rules depend on the habitual environmental and social parameters controlling the reliability of private information and the typical alignment of preferences to which individuals are adapted. I then investigate the collective consequences of those individual interactions in groups of varying sizes and show how the characteristics and efficacy of collective decisions change when the connectivity of the group is varied over timescales that do not allow for full adaptation, with a discussion of the implications of these results for both animal and human communities.

\section*{Model}
I consider a group of $n$ agents sequentially choosing between two options, A and B. The true utilities of A and B are unknown by the agents and may differ between agents; for agent $i$ these are denoted as $U_{A, i}$ and $U_{B, i}$, and I define the difference between these utilities as $x_i \equiv U_{A,i}-U_{B,i}$ \cite{mann2018cdm, mann2020collective}. Agents are assumed to be rational decision makers who seek to maximise the expected utility of their decision \cite{von2007theory} based on the total information $I$ that they have.
\begin{equation}
    P(i \rightarrow \textrm{A}) = 
    \begin{cases}
    1 \ \textrm{if} \ \mathbb{E}(x_i \mid I) > 0 \\
    0 \ \textrm {otherwise}
    \end{cases}
\end{equation}
Agents estimate the value of $x_i$ based on both private information received from the environment and social information in the form of observed decisions by other agents. Following previous work \cite{mann2018cdm, mann2020collective, mann2021optimal}, agents start with normally distributed prior beliefs over $x_i$, with zero mean (no prior information in favour of A or B) and unit variance (setting the arbitrary scale of utility):
\begin{equation}
    p(x_i) = \phi(x_i), \label{eqn:prior}
\end{equation}
where $\phi(\cdot)$ is the standard normal probability density function. Alignment between agents' preferences are represented by the covariance of utility differences, defined as $\textrm{cov}(x_k, x_l) = \rho$ for any two different agents $k$ and $l$. 

Each agent receives private information, $\Delta_i$, from the environment that represents a noisy signal of the true value of $x_i$:
\begin{equation}
 p(\Delta_i \mid x_i) = \phi((x_i-\Delta_i)/\epsilon), \label{eqn:privateinfo}
\end{equation}
where $\epsilon$ controls the reliability of this private information. Given social information, $\mathbf{s}$, agent $i$ forms a posterior belief about the value of $x_i$ by combining its prior, its private information and the social information:
\begin{equation}
    p(x_i \mid \Delta_i, \mathbf{s}) \propto p(x_i)p(\Delta_i \mid x_i)P(\mathbf{s} \mid x_i)
\end{equation}
In the context of this model, social information  $\mathbf{s}$ constitutes an ordered sequence of previous decisions observed by the focal agent prior to making its own decision. The effect of social information $\mathbf{s}$ on the decision of agent $i$ is therefore determined by the detailed structure of $P(\mathbf{s} \mid x_i)$ -- the probability that this social information would be observed conditioned on the value of $x_i$. If the observed social information is more probable conditioned on high values of $x_i$, agent $i$ will accordingly adjust its belief in favour of option A and vice versa, weighing this against its private information $\Delta_i$. 

In judging the social information it observes, a rational agent must recognise that this information is potentially incomplete, since not all other individuals may be observable. Those that the focal agent can observe may in turn have been influenced by each other, or by further, non-observed individuals. In this paper I assume that each agent observes any other with a fixed probability $q$, and this set of observation relations thus forms a random directed network that may range from fully-connected ($q=1$), where all agents observe each other, to sparse ($q \ll 1$), where each agent only observes a small subset of the overall group. How an agent uses the social information it can observe thus depends on that agent's beliefs about the network connectivity. I assume that in any given group agents are habituated to a specific social connectivity $q_{\textrm{habitual}}$, whether by experience or evolutionary adaptation, under which their behaviour is optimised. That is, they act so as to maximise their expected utility, with that expectation being conditioned social connectivity remaining at this habitual value. This may differ from the true connectivity at the time behaviour is observed, $q_{\textrm{actual}}$, as a result of social or environmental changes that alter the connectivity more quickly than adaptation can occur, leading to potential changes in behaviour that may not be optimal for the new context.

An optimal decision-making rule within this model is one that provides a unique decision (A or B) for any combination of private information ($\Delta_i$) and social information ($\mathbf{s}$), such that the individual maximises the expected utility of their choice. That is:
\begin{equation}
    P(i \rightarrow \textrm{A}) = 
    \begin{cases}
    1 \ \textrm{if} \ \int_{-\infty}^\infty x_i p(x_i)p(\Delta_i \mid x_i)P(\mathbf{s} \mid x_i)dx_i > 0 \\
    0 \ \textrm {otherwise}
    \end{cases}
    \label{eqn:utility_max} 
\end{equation}

Equation \ref{eqn:utility_max} implies that there is a critical threshold $\Delta_{\mathbf{s}}^*$, associated with any given social information, such that agent $i$ will choose A if and only if $\Delta_i > \Delta_{\mathbf{s}}^*$. From the perspective of a rational agent, decisions are fully determined by the value of their private information relative to the appropriate threshold specified by the social information they observe. However, when considering the response of an agent to social information, the private information they observe is unknown. This results in uncertainty in the decision the agent will make conditioned on a given set of social information $\mathbf{s}$, such that the response to social information can be characterised by an apparently stochastic decision rule:
\begin{equation}
    P(i \rightarrow \textrm{A} \mid \mathbf{s}) = P(\Delta_i > \Delta_{\mathbf{s}}^*).
\end{equation}
More details on the mathematical structure and derivation of optimal decision-making rules is given in Methods.

\section*{Results}

\subsection*{Effect of sparse connectivity on optimal social response}
\begin{figure*}[h!]
\centering
\includegraphics[width=17cm]{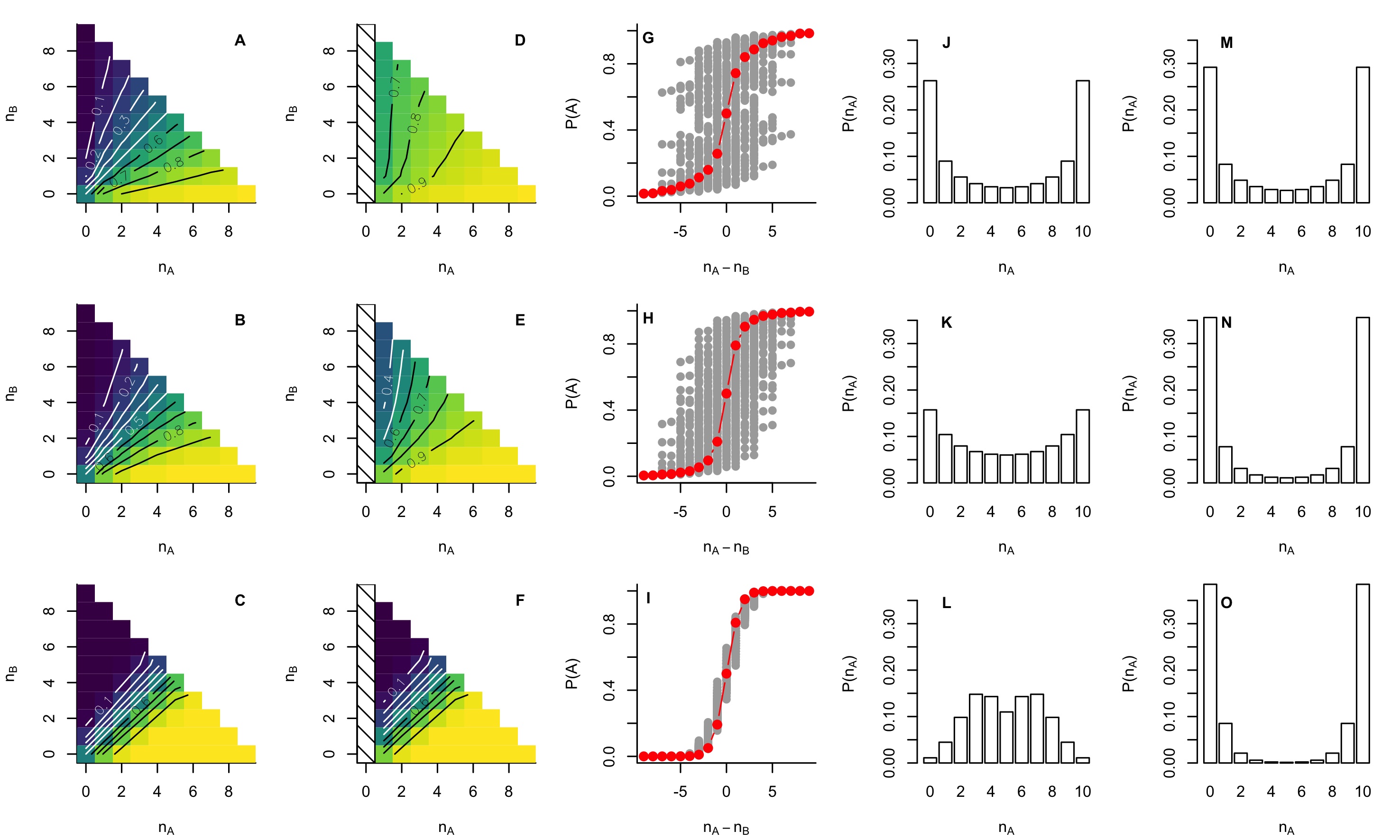}
\caption{Characterising the optimal social information strategy for differing values of social connectivity. Each row shows differing perspectives on the social response by agents habituated to a specific level of social connectivity: $q_{\textrm{habitual}}=0.9$ (top row), $q_{\textrm{habitual}}=0.5$ (middle row) or $q_{\textrm{habitual}}=0.1$ (bottom row). Panels A-C show the probability a focal agent will choose option A, conditioned on the number of previous decisions for A and B observed by that agent. Panels D-F show the same, but further conditioned on the most recently observed decision being for A (thus also ruling out instances where $n_A=0$). Panels G-I show the social response projected onto one dimension: the difference $n_A-n_B$; grey points indicate probabilities arising from specific sequences, while red points show the mean probability for each value of $n_A-n_B$. Panels J-L show the consequence of these decision rules in terms of the distribution of collective outcomes (total number of decisions for A) under habitual conditions ($q_{\textrm{actual}}=q_{\textrm{habitual}}$, where social connectivity remains at the level the agents are adapted to), while panels M-O show the distribution of collective outcomes under conditions of full connectivity ($q_{\textrm{actual}}=1$) if agents continue to apply the behavioural rules optimised for their habitual conditions.}
\label{fig:smallgroup}
\end{figure*}

I first derived in full the optimal decision-making rule for agents in a group of ten identical individuals ($\rho=1$), sequentially choosing between two options A and B, with varying degrees of habitual connectivity, $q_{\textrm{habitual}}$. Specifically I consider cases where agents observe 90\% of other decisions ($q_{\textrm{habitual}}=0.9$), 50\% ($q_{\textrm{habitual}}=0.5$) and 10\% ($q_{\textrm{habitual}}=0.1$) on average. To characterise the optimal behaviour I mapped the probability that a focal agent will select option A conditioned on the number of previous individuals that it observes to have chosen A or B ($n_A$ and $n_B$ respectively), assuming that (unknown to the decision-maker) the two options are in fact identical ($x=0$). This probability is shown for the three cases in Figure \ref{fig:smallgroup} panels A ($q_{\textrm{habitual}}=0.9$), B ($q_{\textrm{habitual}}=0.5$) and C ($q_{\textrm{habitual}}=0.1$). Panel A (high connectivity) shows a pattern of social response very similar to the previously derived case of full connectivity \cite{mann2018cdm}. In panel B (intermediate connectivity) the social response is stronger (greater probability to follow the majority) and in panel C (low connectivity) stronger still -- agents adapt their strategy to attend more strongly to social information when it is limited. As the connectivity is decreased the contour lines of equal probability also transform from a radial pattern (panel A), which maps to a proportional difference between $n_A$ and $n_B$, to a parallel pattern (panel C) that more closely maps to an absolute difference. 

More insight into the form of the social response is gained by conditioning on those cases in which the most recent observable decision-maker selected option A, shown in panels D-F. In the case of high connectivity (panel D) this conditioning makes a large difference to the probability for the focal agent to choose A. As connectivity is decreased the effect of this most recent decision diminishes, and in the case of sparse connectivity is negligible (panel F), showing that the agents in this case respond only to aggregate quantities ($n_A$ and $n_B$) and effectively ignore the ordering of previous decisions. 

Panels G-I show the social response projected onto a single dimension: the absolute difference $n_A-n_B$. Grey points indicate the probability conditioned on each potential sequence of past decisions, while red points indicate the average of these (weighted by the probability of the sequence arising) for each possible value of the difference. This shows that the initially sequence-dependent response in panel G transforms to depend only on the quantity $n_A-n_B$ when the connectivity is lowered (panels H-I).

The collective consequences of these decision rules can be quantified by the total number of agents that ultimately select option A. Here one can distinguish between two possible scenarios. The first is a collective decision made under the conditions of social connectivity the agents are optimised for ($q_{\textrm{actual}}=q_{\textrm{habitual}}$). This is shown in panels J-L, with a tendency towards consensus in the case of high connectivity (panel J), which decreases in line with decreasing connectivity. Although agents optimised for low social connectivity have a stronger response to the other agents they can observe (see above), this does not compensate for the overall reduction in social information, so agents operating with low connectivity are much less likely to all select the same option (panels K-L). This pattern reverses, however, if the decision is made under a different regime, where all agents can observe each other ($q_{\textrm{actual}}=1$), but continue to apply the behavioural rules that are optimised for their habitual conditions. As shown in panels M-O, under this treatment agents optimised for conditions of low connectivity are more social, resulting in a greater degree of group consensus. Such a change in social connectivity might result from, for example, translating animals into a laboratory environment that lacks the usual sources of sensory occlusion that the wild habitat contains, or by placing human subjects in a computer-mediated environment that reveals the decisions of all group members in a way that would not occur in more naturalistic settings. 

\subsection*{Behaviour of large, sparsely-connected groups}
\begin{figure}[h!]
\centering
\includegraphics[width=12cm]{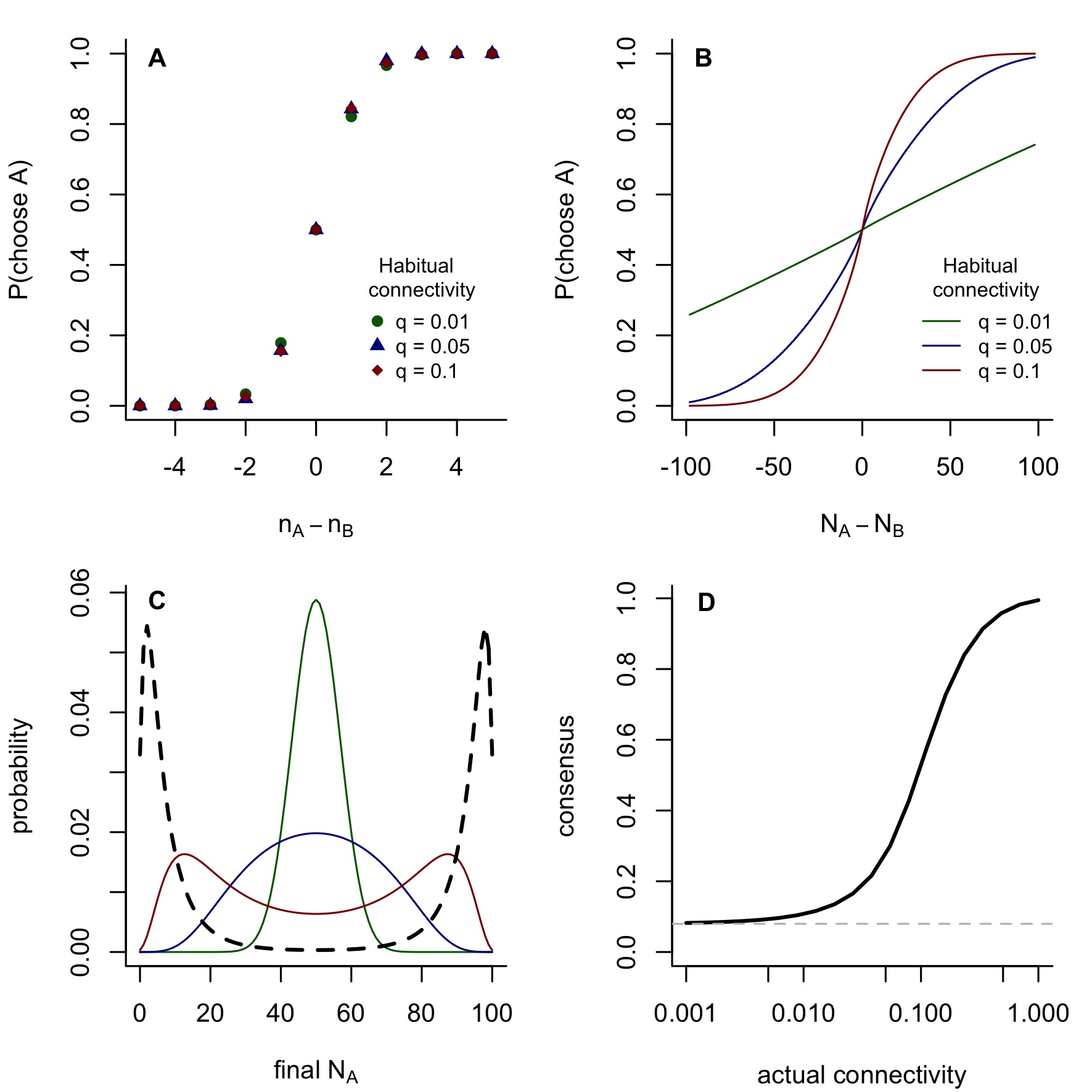}
\caption{Characterising the optimal social information strategy in a large ($n = 100$), sparsely-connected network, for agents habituated to differing connectivity ($q_{\textrm{habitual}} = \{0.01, 0.05, 0.1\}$). (A) Agents adopt a decision rule that depends on the observed difference in the number of prior decisions between options ($n_A-n_B$). This rule is essentially independent of the connectivity to which they are habituated. (B) When this rule is applied under the habitual conditions, the observable response to the true difference ($N_A-N_B$) is weaker in conditions of lower connectivity; (C) More sparsely-connected groups are less cohesive under natural conditions (solid lines), as measured by the total number of agents choosing one option. Increasing network connectivity to $q_{\textrm{actual}}=0.2$ and applying the same decision rule results in a substantially more cohesive collective outcome that is independent of the habitual connectivity (dashed line). (D) The degree of consensus ($|N_A-N_B|/(N_A+N_B)$) varies monotonically with the true connectivity of the network and is independent of habitual connectivity. The dashed grey line indicates the expected consensus when agents choose independently.}
\label{fig:largegroup}
\end{figure}

To analyse the behaviour of larger group sizes that characterise many animal and human societies one can utilise the emergent property of sparsely-connected groups revealed above: the dependence of the social response solely on the absolute difference $n_A-n_B$. This property means that an optimal decision rule for agents in a group of size $n$ can be characterised by responses to $2n+1$ distinct possible values of $n_A-n_B$, and this can be further reduced by applying the symmetry in response to absolute differences of equal magnitude but differing sign. Using a Monte Carlo method (see Methods), I derived the optimal decision-making rule for agents in a group of 101 individuals, with a habitual connectivity of either $q_{\textrm{habitual}}=0.01$, $q_{\textrm{habitual}}=0.05$ or $q_{\textrm{habitual}}=0.1$ (equivalent to a mean degree of 1, 5 or 10). This optimal strategy is illustrated in Figure \ref{fig:largegroup}A, which shows that the probability for a focal agent to select option A, conditioned on the value of $n_A-n_B$ it observes, is almost identical across the differing levels of habitual connectivity. I then calculated the probability that an agent will select option A, conditioned on the difference $N_A-N_B$, where $N_A$ and $N_B$ refer to the actual number of previous decision makers selecting either A or B, not only those that the focal agent can observe. That is, this probability represents that which would be inferred by making a simple empirical analysis of the proportion of agents choosing option A based on the number of previous decisions for A and B seen by an external observer who can observe all decisions but who does not have access to the network defining which agents can observe each other. This probability, shown in Figure \ref{fig:largegroup}B, varies strongly with habitual connectivity. Taken together, the results in panels A and B indicate that agents in more densely-connected (but still sparse) networks exhibit more sociality by virtue of increased social connections, rather than as a result of any difference in how they use that social information. To see how this is reflected in collective outcomes, I calculated the probability that a given proportion of agents will ultimately select option A under each level of habitual connectivity, under habitual conditions ($q_{\textrm{actual}}=q_{\textrm{habitual}}$). The results in Figure \ref{fig:largegroup}C show that agents in the more connected network ($q_{\textrm{habitual}}=0.1$, red line) exhibit a tendency to consensus, with the peaks of the distribution near 0 and 100, while those in the less connected networks show a distribution that is peaked at 50, indicating a tendency for the group to split evenly between the options; when $q_{\textrm{habitual}}=0.01$ this is barely distinguishable from a binomial distribution in which each agent chooses independently (blue line). However, the same three sets of agents, transferred to a new condition with increased connectivity of $q_{\textrm{actual}}=0.2$ (mean degree = 20), all exhibit identical collective behaviour, with a strong bias towards choosing the same option (dashed black line). Finally, I repeated this analysis for actual connectivity levels from $q_{\textrm{actual}}=0.01$ (mean degree = 1) to $q_{\textrm{actual}}=1$ (mean degree = 100) and measured the resulting consensus, defined as $|N_A-N_B|/(N_A+N_B)$, once all decisions have been made. As shown in Figure \ref{fig:largegroup}D, group consensus varies strongly and monotonically with $q_{\textrm{actual}}$.

\subsection*{Effect of differing preferences}
The binary decision problem studied here was also considered in ref. \cite{perez2011cab}. There the authors present a derivation for a decision rule based on an approximation under which other agents are assumed to have made their choices independently. In the case of a symmetric choice, they arrive at a decision rule of the form:
\begin{equation}
    P(i \rightarrow \textrm{A} \mid n_A, n_B) = \frac{1}{1+S^{-(n_A-n_B)}}, \label{eqn:alfonso}
\end{equation}
where $S$ is a sociality parameter that represents the relative likelihood that another individual will make a `correct' or `incorrect' choice. Since the effect of social information in this rule also depends on the value of $n_A-n_B$ it is natural to compare how that study relates to this one. In the case of a symmetric choice ($x_i = 0 \ \forall i$), rearranging equation \ref{eqn:alfonso} suggests an effective value of the sociality parameter:
\begin{equation}
        S = \left(1/P(i \rightarrow \textrm{A} \mid n_A-n_B =1) -1 \right)^{-1} \label{eqn:effS}
\end{equation}
where $P(i \rightarrow \textrm{A} \mid n_A-n_B =1)$ is the probability that agent $i$ will choose option A, having observed social information $n_A-n_B=1$ (see Methods).

\begin{figure}[h!]
\centering
\includegraphics[width=12cm]{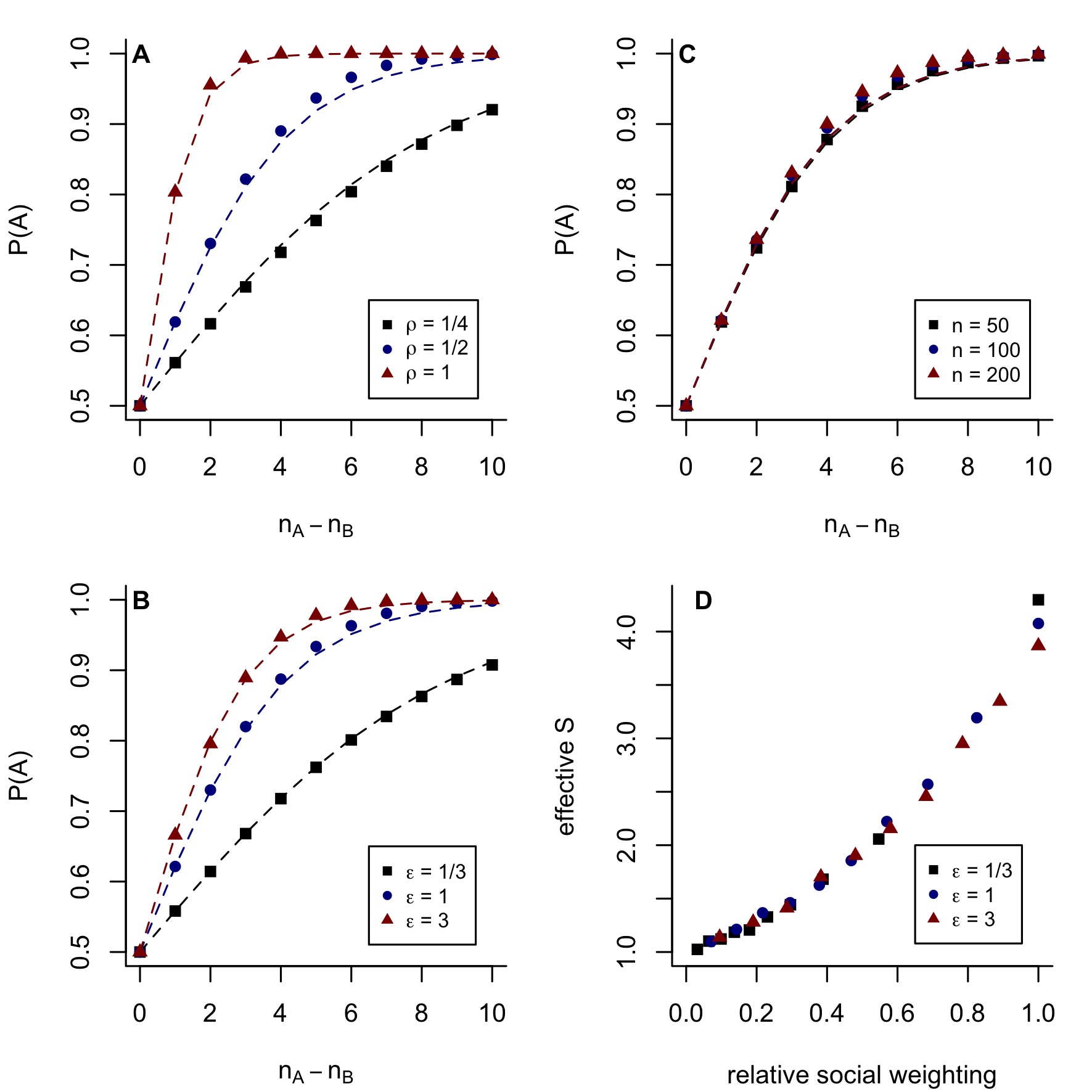}
\caption{Variation of optimal social responses with differing preference alignment $\rho$, private information variance $\epsilon$ and group size $n$. (A) variable $\rho$, with fixed $\epsilon=1$, $n=100$; (B) variable $\epsilon$ with fixed $\rho=0.5$ and $n=100$; (C) variable $n$ with fixed $\rho=0.5$ and $\epsilon =1$. In each case is also plotted the probabilities given by equation \ref{eqn:alfonso} (dashed lines), with sociality parameter $S$ given by equation \ref{eqn:effS}, demonstrating a close fit between the two formulations across widely varying parameters; (D) the calculated value of $S$ in a group of size $n=100$ as a function of the `relative social weighting' (RSW), for varying values $\rho$ and $\epsilon$.}
\label{fig:alfonso}
\end{figure}

In Figure \ref{fig:alfonso} I show that equation \ref{eqn:alfonso} provides a highly accurate approximation to the probability that a focal agent will choose option A in a symmetric setup, with an effective value of $S$ that increases strongly with the degree of correlation between agent preferences ($\rho$, panel A) and the variance of private information ($\epsilon$, panel B), and weakly with group size ($n$, panel C). In ref. \cite{mann2020collective}, a combination of $\rho$ and $\epsilon$ termed the `relative social weighting' ($\textrm{RSW} = \epsilon \rho / \sqrt{1+\epsilon^2-\rho}$) was found to be highly predictive of social behaviour; here the same combination is seen to determine the effective value of $S$, again controlling the magnitude of social response (panel D).

There is, however, a key difference between the this study and ref. \cite{perez2011cab}. Rather than deriving an approximate decision rule by assuming that agents treat their observations of others' choices as being statistically independent (as in ref. \cite{perez2011cab}), here this quasi-independence emerges from the sparsity of the interaction network. Note also that while these observed decisions are treated as being independent of each other, this need not imply they are independent of any social information, only that the social information they in turn depend on is not observed by the focal agent. As a corollary, this difference makes it clear that the $n_A$ and $n_B$ that the focal agent responds to are the number of decisions for A and B that the agent itself perceives, which may be different to those recorded by, say, an experimental observer (cf. Figure \ref{fig:largegroup}A and B). Thus, statistical efforts to fit equation \ref{eqn:alfonso} to experimental or observational data (see e.g \cite{farine2013, eguiluz2015bayesian} as well as examples in \cite{perez2011cab}) are valid when the agents under observation habitually inhabit a sparse interaction network, and insofar as the recorded social information accurately represents the individuals the focal agent actually attended to. Otherwise, while the basic sigmoidal form of equation \ref{eqn:alfonso} may provide an acceptable description of the observed decision probabilities, the inferred parameter values will not represent true properties of the decision-making process.

\subsection*{Impact of changing connectivity on individual rewards}
The results above show that agents adapted to sparse social connections will exhibit stronger social responses when connectivity is increased. What impact will this have on the rewards the agents are able to extract from the environment? Naively one might expect that any deviation from the condition they agents are adapted to will be detrimental, but that expectation may be deceptive in this case;  social connectivity increases the total social information available to each agent, and the benefit of this additional information may outweigh the costs of being adapted to a different social environment.

\begin{figure}[h!]
\centering
\includegraphics[width=12cm]{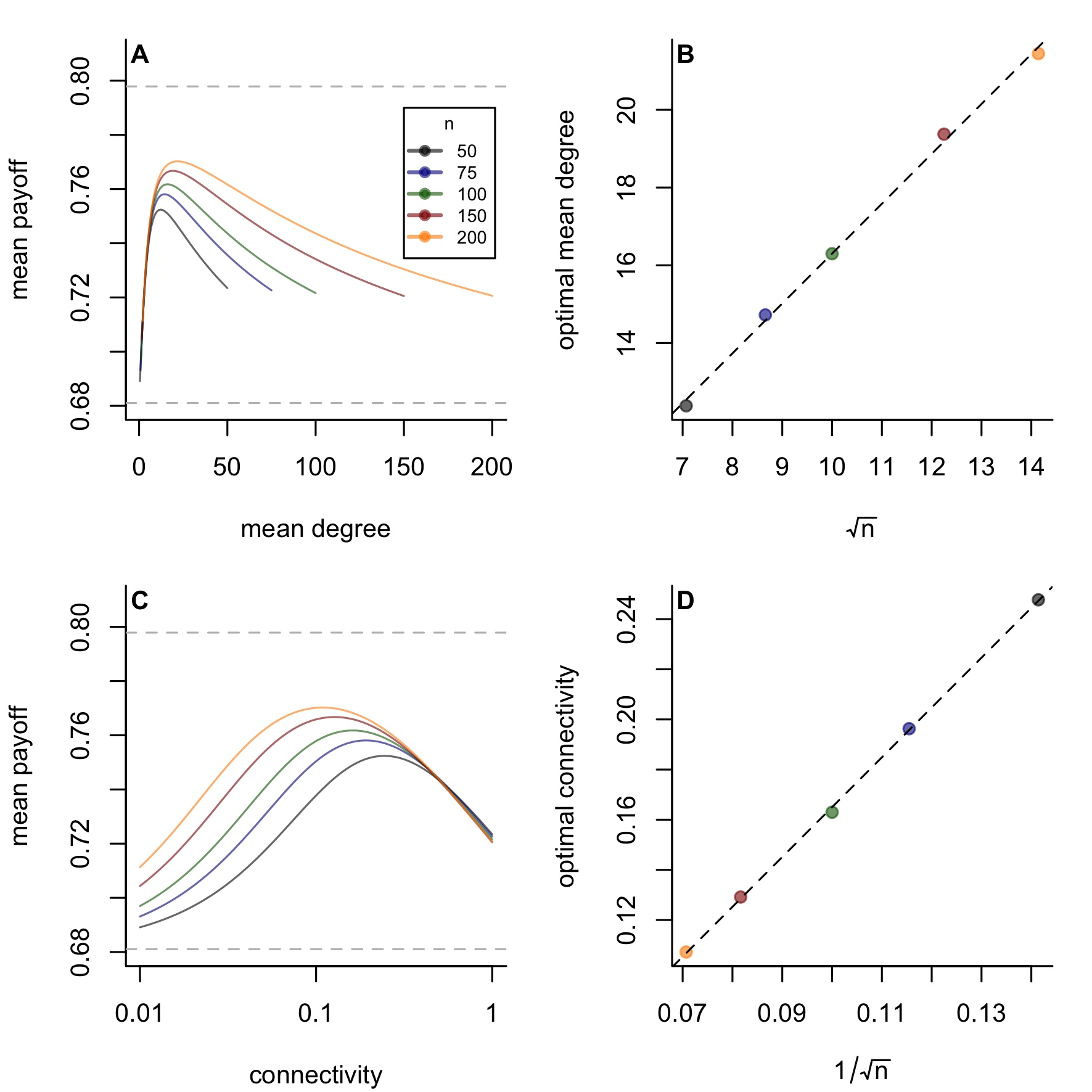}
\caption{Average payoff obtained by agents in sparse networks of varying sizes as a function of actual connectivity and mean degree. Agents in each network are habituated to a connectivity $q_{\textrm{habitual}} = 5/(n-1)$ implying a mean degree of 5, though the results are not sensitive to this factor as long as habitual connectivity is sparse. (A) Mean payoffs increase as mean degree increases in all networks up to an optimal value, and then fall, with a differing optimal mean degree for different group sizes. The dashed grey lines indicate the average reward for a single agent without any social information (lower line) and the average reward for an agent that always chooses the option with greater utility (upper line); (B) The optimal mean degree generating maximum payoffs varies proportionally to the square root of network size; (C) Mean payoffs increase with connectivity up to an optimal value and then fall, with a differing optimal connectivity for different group sizes; (D) The optimal connectivity varies inversely with the square root of network size.}
\label{fig:rewards}
\end{figure}

I calculated the expected payoff per decision for identical agents ($\rho=1)$ adapted to sparse connectivity (mean degree = 5) in groups of varying size between $n=50$ and $n=200$ under a range of experimental social connectivities between $q_{\textrm{actual}}=0.01$ and $q_{\textrm{actual}}=1$. This expected payoff is plotted in Figure \ref{fig:rewards} as a function of mean degree (panel A) and connectivity (panel B). These results show that for any group size there is an optimal connectivity, at which the expected payoff is maximised. This optimal connectivity varies consistently with group size, such that the optimal mean degree varies proportionally to the square root of the group size (panel C), and equivalently the optimal value of $q_{\textrm{actual}}$ varies in inverse proportion to $\sqrt{n}$ (panel D). Above this optimal value, further increases in connectivity reduce expected payoffs.

\section*{Discussion}
In this paper I derived the optimal use of social information arising from incomplete observations of the decisions of other agents. I showed that when such information is sufficiently sparse, rational agents develop a strategy that attends only to the absolute difference between the number of other individuals that have selected each option, ignoring both the total number of previous decisions and the order in which they were made. Exploiting this property to analyse large groups revealed that the optimal social response is essentially independent of the degree of social connectivity the agents are habituated to; behaviour is instead controlled by the connectivity at the time the decision is made. In particular, greater connectivity leads to increasing consensus among group members. Finally I showed that increases in social connectivity are beneficial to sparsely-connected groups up to a critical value of $q$, increasing the average payoff of the options individuals choose. However, sufficiently large increases in connectivity result in extreme social feedback that ultimately leads to lower average rewards.

This model predicts that the number of social connections individuals experience at the time they make a decision determines the degree of group consensus. This has practical consequences for the scientific study of collective behaviour. For example, an animal that lives in conditions of high sensory occlusion such as muddy water may respond weakly to local con-specifics simply because it does not perceive many of them; if translated to an environment with less occlusion such animals may exhibit increased apparent social responses without fundamentally changing how they respond to the other individuals that they can perceive. While in experiments one might expect animal or human subjects to adjust somewhat to the new context (and indeed animal subjects are often given time in the new environment to acclimatise), this finding provides an important context for laboratory-based research, especially since the effect operates in the same direction as earlier findings that decreased noise in environmental signals could also increase observed sociality in laboratory conditions \cite{mann2018cdm}. 

Where rational individuals experience sparse social connectivity they are able to treat the decisions of others as being effectively independent. As long as this assumption remains valid, more social information, in the form of additional social connections, is beneficial to a decision-maker, resulting in a greater average payoff from their decisions. However, if social connectivity exceeds a critical value this independence assumption begins to breakdown, and if agents do not recognise and adjust for this they will make worse decisions. In this model, this critical value of social connectivity was found to depend on the square root of the total group size, such that $q_{\textrm{optimal}} \propto 1/\sqrt{n}$, giving a critical mean degree proportional to $\sqrt{n}$. This scaling resembles that found in previous studies of optimal network structure for collective decision-making \cite{kao2019modular}, suggesting a common basis in agents applying decision rules that assume independence between those others who they observe.

The relationship between connectivity and decision-making efficacy has profound implications for the welfare and sustainability of animal and human communities. Under pressure from climate change and habitat destruction, many animal populations have experienced precipitous declines. Many of the affected populations are social animals, and these may depend on social information, for instance to locate food sources or navigate to seasonal habitats. Reduction in group size while holding constant the probability of social interaction between any pair of individuals is likely to lead to individuals making poorer decisions on average (see Figure \ref{fig:rewards}C), potentially creating additional difficulties for populations already under pressure. There is evidence that in some cases animals may seek to compensate for lost social bonds by creating new connections \cite{firth2017wild}, but this may depend on whether populations are adapted to significant changes in population size over time. These results suggest therefore that, where populations are in decline, maintaining the physical cohesion and density of animal groups, so that social connectivity remains at appropriate levels, should be an important conservation goal alongside wider efforts to promote landscape connectivity \cite{rudnick2012role, doherty2018coupling}. 
 
In human societies a more salient concern is the efficacy of decision making by individuals exposed to recent increases in social connectivity \cite{bak2021stewardship}, most obviously through social media \cite{kietzmann2011social}, but also through the ubiquity on online recommendation and reputation systems \cite{dellarocas2003digitization}. These factors serve to increase the number of choices, decisions and opinions individuals can observe others making or expressing. The rapidity with which this social environment has changed for many people, and the time required to adapt to novel social contexts \cite{burton2021payoff}, makes it highly plausible that they will continue to operate with heuristics that are still adapted to lower levels of connectivity. While some increase in connectivity may be beneficial, large increases have the potential to reduce the quality of decision making by all participants by introducing a high degree of social feedback that swamps the true quality signals from the environment rather than amplifying them, such that collective decisions become highly contingent on variable early decision makers \cite{Salganik:2006mi, muchnik2013social}. A number of experimental studies have revealed reductions in collective wisdom when individuals are exposed to the complete or aggregated opinions of other group members \cite{lorenz2011hsi, frey2020social}, but others have shown improvements under treatments where connectivity is limited \cite{kao2014decision,becker2017network, navajas2018aggregated,kao2019modular, jayles2020exchanging}. These apparently contrary findings may be reconciled by recognising that some additional communication between agents is beneficial, but that too much can be harmful when the opinions or decisions of agents become highly correlated in manner not recognised or accounted for by another agent observing them. 

The potential endogenous reduction in decision-making efficacy identified above has political implications in an age of ubiquitous online connectivity, but should also be a source of concern in an increasingly-connected scientific community \cite{mann2017maintaining}. Societies that have recently experienced large increases in connectivity may also be more vulnerable to exogenous misinformation (e.g. \cite{spaiser2017twitter}), as false information can be more easily amplified throughout the population. In social networks that are naturally highly-connected, such as small friendship groups individuals may intuit the correlations arising from those connections and apply appropriate decision rules \cite{mann2018cdm, mann2020collective, mann2021optimal}. However, such correlations may now be induced in wider networks where they were previously absent as a result of increased connectivity, with the information nonetheless appearing to come from independent sources to an agent that has not adjusted to the new social context. It is these cases where habituation and expectations deviate from reality that are most ripe for the spread of `viral' opinions. This should motivate efforts not only to identify and remove obvious misinformation online, but also to educate citizens in how to respond to social information in a highly-connected world \cite{badrinathan2021educative, bak2021combining}.

\section*{Methods}
\subsection*{Correlation between agent preferences}
The decisions of other individuals are useful information insofar as they are assumed to share preferences with the focal agent. Following \cite{mann2020collective}, I encode the degree to which individuals' preferences align via a joint multivariate normal distribution $\mathcal{N}(\cdot, \cdot, \cdot)$:
\begin{equation}
    p(x_1, x_2, \ldots x_n) = \mathcal{N}\left(\mathbf{x}, \mathbf{0}, \mathbf{\Sigma} \right). \label{eqn:U_mvn}
\end{equation}
where $\mathbf{x} = [x_1, \ldots, x_n]^\top$, $\mathbf{0}$ is vector of $n$ zeroes, and $$\mathbf{\Sigma} = \begin{bmatrix} 1 & \rho &\ldots & \rho \\ \rho & \ddots & & \vdots \\ \vdots &  &  & \rho \\ \rho & \ldots & \rho & 1 \end{bmatrix},$$ is the covariance matrix, with $\rho$ being the correlation between the utilities of any pair of agents. Combined with equation \ref{eqn:privateinfo} this further specifies a joint multivariate normal distribution over the true values of $x_1, \ldots x_n$ and the observed private information of each individual $\Delta_1, \ldots \Delta_n$, where:
\begin{equation}
\begin{split}
   & \textrm{cov}(x_k, x_l) = \textrm{cov}(\Delta_k, x_l) = \rho + \delta_{k, l}(1-\rho)\\
& \textrm{cov}(\Delta_k, \Delta_l)= \rho + \delta_{k, l}(1+\epsilon^2-\rho),
\end{split}
\end{equation}
where $\delta_{k,l}$ is the Kronecker delta function. A key consequence of this joint distribution is the probability of observing $\Delta_i$ conditioned on the true value of $x_j$:
\begin{equation}
    p(\Delta_j \mid x_i) \propto \phi\left(\frac{\rho x_i - \Delta_j}{\sqrt{1+\epsilon^2-\rho}} \right),
\end{equation}
and thus the probability that $\Delta_j$ exceeds some threshold $\Delta^*$, conditioned on $x_i$ is given by:
\begin{equation}
    p(\Delta_j > \Delta^* \mid x_i) =\Phi\left(\frac{\rho x_i - \Delta^*}{\sqrt{1+\epsilon^2-\rho}} \right),
\end{equation}
where $\Phi(\cdot)$ is the standard cumulative normal distribution function.

\subsection*{Form of social information strategy}
Based on the methodological developments in previous studies \cite{mann2018cdm, mann2020collective, mann2021optimal}, a social information strategy associates a real number, $\Delta^*_{\mathbf{s}}$ with each distinct observable state of social information $\mathbf{s}$. This number represents a threshold, such that if an agent receives private information $\Delta > \Delta^*_{\mathbf{s}}$ the agent will choose option A, otherwise it will choose option B. 

In an optimal strategy the thresholds $\Delta^*_{\mathbf{s}}$ are identified as those that cause the agent to make rational decisions, i.e. to choose the option that maximises its expected utility, conditioned on both the private and social information it has available. This implies that when an agent observes private information exactly equal to the threshold value, this makes the expected utility difference between the two options ($x_i$) equal to zero:
\begin{equation}
\begin{split}
    0 &= \mathbb{E}(x_i \mid \mathbf{s}, \Delta_i=\Delta^*_{\mathbf{s}})\\ &\propto \int_{-\infty}^\infty x_ip(x_i)p(\Delta_i = \Delta^*_{\mathbf{s}} \mid x_i)P(\mathbf{s} \mid x_i)dx_i \label{eqn:basicform}
    \end{split}
\end{equation}
In this paper the distinct observable states of social information are observed sequences of decisions, $\mathbf{s} = K_1, \ldots K_k$, which are themselves random subsets of the true full sequence of decisions $\mathbf{c} = C_1, C_2, \ldots, C_m$, such that $k \leq m$. Given an observed decision sequence $\mathbf{s}$, equation \ref{eqn:basicform} can be expanded as below:
\begin{equation}
\begin{split}
    0 &= \mathbb{E}(x_i \mid \mathbf{s}, \Delta_i=\Delta^*_{\mathbf{s}}) \\
    &\propto \int_{-\infty}^\infty x_ip(x_i)p(\Delta_i = \Delta^*_{\mathbf{s}} \mid x_i)\sum_{\mathbf{c} \in \mathcal{S}} P(\mathbf{s} \mid \mathbf{c})P(\mathbf{c} \mid x_i)dx_i \label{eqn:expandedform}
\end{split}
\end{equation}
where the summation is over the set $\mathcal{S}$ of all possible sequences of $n-1$ decisions that could have occurred prior to the focal agents decision. 

\subsection*{Observation probability of a decision sequence}
Given a true sequence of previous choices, $\mathbf{c} = C_1, C_2, \ldots, C_m$ and a connectivity parameter $q$ (the probability to observe any given prior decision), the probability that the focal agent will observe the sequence $\mathbf{s} = K_1, \ldots K_k$ is given by:
\begin{equation}
    P(\mathbf{s} \mid \mathbf{c}) = q^k (1-q)^{m-k} \mathcal{C}(\mathbf{s}, \mathbf{c})
\end{equation}
where $\mathcal{C}(\mathbf{s}, \mathbf{c})$ is the number of distinct decimations of $\mathbf{c}$ that can result in $\mathbf{s}$. To give an example, if the original sequence of decisions is $\mathbf{c} = 1, 1, -1, 1, -1$, the observed sequence $\mathbf{s} = 1,1$ can be generated from this in three distinct ways: by selecting elements 1 and 2, elements 1 and 4 or elements 2 and 4. In contrast the observed sequence $\mathbf{s}' = -1,-1$ can only be generated in one way, by selecting elements 3 and 5. Hence, in this example, $\mathcal{C}(\mathbf{s}, \mathbf{c}) = 3$ and $\mathcal{C}(\mathbf{s}', \mathbf{c}) = 1$. By extension then, $P(\mathbf{s} \mid \mathbf{c})= 3q^2(1-q)^3$, and $P(\mathbf{s}' \mid \mathbf{c})= q^2(1-q)^3$. 

\subsection*{Generating probability of a decision sequence}
The summation in equation \ref{eqn:expandedform} requires calculating the probability $P(\mathbf{c} \mid x_i)$ to generate any possible sequence of decisions, conditioned on a specific true value of $x_i$. This quantity can be evaluated by considering the various contiguous sub-sequences of $\mathbf{c}$: $\mathbf{c}_k = C_1, \ldots C_k$. Since the form of the social information strategy specified in equation \ref{eqn:expandedform} specifies a threshold for each of these sub-sequences, the probability of the full sequence is simply the product of the probability for each decision within that sequence, conditioned on those thresholds. The probability for each decision takes the form:
\begin{equation}
\begin{split}
    P(j \rightarrow \textrm{A} \mid x_i, \mathbf{c}) &= \sum_{\mathbf{s} \in \mathcal {S}} P(\Delta_j > \Delta^*_{\mathbf{s}} \mid x_i)P(\mathbf{s} \mid \mathbf{c}) \\
    &= \sum_{\mathbf{s} \in \mathcal {S}} \Phi \left(\frac{(\rho x_i- \Delta^*_{\mathbf{s}})}{\sqrt{1+\epsilon^2-\rho}}\right)P(\mathbf{s} \mid \mathbf{c}),
    \label{eqn:generation}
\end{split}
\end{equation}
where the summation is over all the sequences $\mathbf{s} \in \mathcal {S}$ that the agent may have observed as a result of the true sequence $\mathbf{c}$.

Applied recursively, this relation specifies the probability of the full sequence $\mathbf{c}$, conditioned on the true value of $x_i$:
\begin{equation}
    P(\mathbf{c} \mid x_i) = \prod_{k=1}^{|\mathbf{c}|} \sum_{\mathbf{s} \in \mathcal {S}}\Phi \left(\frac{C_k(\rho x_i- \Delta^*_{\mathbf{s}})}{\sqrt{1+\epsilon^2-\rho}}\right)P(\mathbf{s} \mid \mathbf{c}_{k-1}), \label{eqn:sequence_prob}
\end{equation}
where $|\mathbf{c}|$ is the length (number of decisions) of the sequence $\mathbf{c}$.

\subsection*{Iterative calculation of thresholds}
The formulation above poses a problem, since determining the value of the threshold $\Delta^*_ {\mathbf{s}}$ in equation \ref{eqn:expandedform} requires knowing the thresholds associated with all sequences in $\mathcal{S}$, including $\Delta^*_ {\mathbf{s}}$ itself. This suggests an iterative approach. First, all thresholds are set to an initial value of zero: $\Delta^*_\mathbf{s} = 0$ for all possible social information $\mathbf{s}$. These values used these initial values to calculate the sequence-generation probabilities $P(\mathbf{c} \mid x_i)$ (for any given agent identity $i$), by applying equation \ref{eqn:generation}. Each threshold is then  updated by solving equation \ref{eqn:expandedform} for each $\Delta^*_ {\mathbf{s}}$, based on these sequence probabilities. These steps are then repeated, iterating until the thresholds reach a stable value that provides a self-consistent solution to equation \ref{eqn:expandedform} for all sequences.

\subsection*{Monte Carlo methods for large groups}
On the basis of the results from small groups, I assume that in sparsely connected groups agents employ a social information strategy that depends on the difference between the number of agents they observe to have chosen A and B. Thus, this strategy associates a threshold $\Delta^*_{w}$ with every value $w = n_A-n_B$ that the agent may observe as the difference between the observed number of decisions for A ($n_A$) and the observed number of decisions for B ($n_B$)

Given a true sequence of decisions $\mathbf{s}$ that contains $N_A$ agents choosing option A and $N_B$ choosing B, the probability that the focal agent observes $n_A$ and $n_B$ decisions for each option is given by two independent binomial distributions:
\begin{equation}
\begin{split}
    p(n_A \mid N_A) &= \binom{N_A}{n_A}q^{n_A}(1-q)^{(N_A-n_A)} \\
    p(n_B \mid N_B) &= \binom{N_B}{n_B}q^{n_B}(1-q)^{(N_B-n_B)}.
\end{split}
\end{equation}
The resulting distribution for the difference $w = n_A-n_B$ observed by the focal agent has no closed form, but can be evaluated numerically by summation over the values of $n_A$ and $n_B$
\begin{equation}
    P(w \mid N_A, N_B) = \sum_{n_A = 1}^{N_A}\sum_{n_B = 1}^{N_B}p(n_A \mid N_A)p(n_B \mid N_B)\delta_{(n_A-n_B), w} \label{eqn:binomdiff}
\end{equation}
where $\delta_{(n_A-n_B), w}$ is the Kronecker delta function.

In large groups, exhaustively calculating the probability to generate every possible sequence of decisions is infeasible. Instead, I employ a Monte Carlo methodology. As in the case of small groups, I initialise all thresholds to zero and based on these thresholds I simulate $N$ candidate decision sequences, with random lengths drawn uniformly between 0 and $n-1$ (since the focal agent, observing only the difference $w$, does not attend to how many previous decisions have been made), and based on values of $x$ drawn randomly from the normal distribution specified in equation \ref{eqn:prior}. Extracting the values of $N_A$ and $N_B$ from the generated sequences, and retaining the samples of $x_i$ used to generate them, these sequences permit the following Monte Carlo approximation to the expected value of $x_i$, conditioned on the observable social information $w$:
\begin{equation}
\begin{split}
    &\mathbb{E}(x_i \mid \Delta_i, w) \propto \int_{-\infty}^{\infty} x_ip(x_i)p(\Delta_i \mid x_i) P(w \mid x_i)dx_i \\
    &\simeq (1/\Lambda) \sum_{\lambda = 1}^\Lambda x_\lambda \phi((x_\lambda - \Delta_i)/\epsilon) P(w \mid N_{A, \lambda}, N_{B, \lambda}),
    \end{split}
\end{equation}
where $\lambda$ indexes the $\Lambda$ Monte Carlo samples, and $P(w \mid N_{A, \lambda}, N_{B, \lambda})$ is given by equation \ref{eqn:binomdiff}. Using this approximation, one can calculate a value for the threshold $\Delta^{\textrm{proposal}}_{w}$ such that:
\begin{equation}
\sum_{\lambda = 1}^\Lambda x_\Lambda \phi((x_\lambda - \Delta^{\textrm{proposal}}_{w})/\epsilon) P(w \mid N_{A, \lambda}, N_{B, \lambda}) = 0
\end{equation}
This value of the threshold, which represents the rational strategy conditioned on all other agents continuing to apply the previously initialised threshold values, acts as a proposed value for the update step. To improve convergence I introduce a learning rate $\alpha$, such that the updated threshold ($\Delta^*_{w, {i+1}}$) is a weighted average of the previous value ($\Delta^*_{w, i}$) and the proposal defined above.
\begin{equation}
   \Delta^*_{w, {i+1}} =  \alpha \Delta^{\textrm{proposal}}_{w} + (1-\alpha) \Delta^*_{w, i}
\end{equation}
In my analyses I use $\Lambda =1000$ Monte Carlo samples, and $\alpha = 0.1$ with 100 iterations to reliably ensure convergence to a stable set of threshold values.

\subsection*{Expected payoff} 
I calculated the expected payoff for an individual as follows: for a given value of utility difference ($x$ - equal for all agents when $\rho=1$) I evaluated the probability that a given number of agents would choose the option with greater utility, based on equation \ref{eqn:sequence_prob} for the probability of a given sequence of decisions and averaging over possible sequences. Agents choosing the higher-reward option are assigned a payoff of $x$, such that the expected reward for a randomly-chosen agent $x$ multiplied by the expected proportion of agents selecting the better option. I then integrated over possible values of $x$ according to the assumed distribution of rewards in the environment, which corresponds to the agents' prior belief about $x$ given in equation \ref{eqn:prior}. 

\subsection*{Acknowledgments}
This work was supported by UK Research and Innovation Future Leaders Fellowship MR/S032525/1.

\bibliographystyle{ieeetr}

\end{document}